\documentclass[aps,prd, notitlepage, onecolumn,superscriptaddress,floatfix,letterpaper,nofootinbib, longbibliography]{revtex4-1}

\usepackage{amsmath, amsthm, amssymb,slashed}

\usepackage[usenames, dvipsnames]{color}
\usepackage[svgnames]{xcolor}
\usepackage[colorlinks,citecolor=RoyalBlue, urlcolor=RoyalBlue, linkcolor=RoyalBlue ]{hyperref} 

\usepackage[normalem]{ulem}

\usepackage{booktabs}

\definecolor{mygray}{gray}{0.6}

\usepackage{upgreek}
\usepackage{bbm}

\newenvironment{myfont}[2][]{\csname#2\endcsname[#1]}{}

\usepackage{slashed}
\usepackage[makeroom]{cancel}
\usepackage[normalem]{ulem}
\usepackage{soul}
\newcommand{\stkout}[1]{\ifmmode\text{\sout{\ensuremath{#1}}}\else\sout{#1}\fi}

\usepackage{sseq}
\usepackage[all,cmtip]{xy}
\usepackage{tikz-cd}
\usepackage{tikz}
\usetikzlibrary{matrix}
\usetikzlibrary{decorations.markings}
\usetikzlibrary{tikzmark,decorations.pathreplacing,positioning}
\usepackage{amsfonts}
\usepackage{multirow}

\newcommand{\bea}{\begin{eqnarray}}
\newcommand{\eea}{\end{eqnarray}}
\def\be{\begin{equation}}
\def\ee{\end{equation}}

\newcommand{\ii}{\hspace{1pt}\mathrm{i}\hspace{1pt}}

\def\CP{{\mathbb{CP}}}

\definecolor{red}{rgb}{1,0,0}
\definecolor{blue}{rgb}{0,0,1}
\definecolor{dblue}{rgb}{0,0,0.4}
\definecolor{green}{rgb}{0,1,0}
\definecolor{black}{rgb}{0,0,0}
\definecolor{white}{rgb}{1,1,1}

\definecolor{brn}{rgb}{.8,.4,.0}
\definecolor{redo}{rgb}{1,.5,.0}
\definecolor{ddgrn}{rgb}{0,0.4,0}
\definecolor{dgrn}{rgb}{0,0.55,0}
\definecolor{dbl}{rgb}{0,0,0.5}

\usepackage[bbgreekl]{mathbbol}
\usepackage{amscd}

\newcommand{\Z}{\mathbb{Z}}
\newcommand{\C}{\mathbb{C}}

\newcommand{\dd}{\mathrm{d}}
\newcommand{\<}{\langle} 
\renewcommand{\>}{\rangle}

\newcommand{\Eq}[1]{Eq.~(\ref{#1})} 
\newcommand{\eq}[1]{eq.~(\ref{#1})} 
\newcommand{\eqq}[1]{(\ref{#1})}

\newcommand{\Tr}{{\rm Tr}}

\newcommand{\prt}{\partial}

\newcommand{\bpm}{\begin{pmatrix}}
\newcommand{\epm}{\end{pmatrix}}
\newcommand{\bmm}{\begin{matrix}}
\newcommand{\emm}{\end{matrix}}

\newcommand{\cG}{ {\cal G} }

\newcommand{\al}{\alpha} 
\newcommand{\bt}{\beta}

\def\Z{{\mathbb{Z}}}

\def\C{{\mathbb{C}}}

\def\Tr{{\mathrm{Tr}}}

\def \Z{\mathbb{Z}}

\def \CP{\mathbb{CP}}

\newcommand {\emptycomment}[1]{}

\newcommand{\U}{{\rm U}}
\newcommand{\SU}{{\rm SU}}

\usepackage{centernot}

\newcommand{\Sec}[1]{Sec.~\ref{#1}} 

\usepackage{enumitem} 
\usepackage{mathtools,amssymb,varwidth}

\newcommand{\Table}[1]{Table \ref{#1}}

\usepackage{datetime}

\def\bB{{\mathbf{B}}}

\newcommand{\rF}{{\rm F}}

\newcommand{\SM}{{\rm SM}}

\def\GCS{\mathrm{GCS}}

\usepackage{array}

\begin{document}


\title{Topological Leptogenesis}

\author{Juven Wang}
\affiliation{London Institute for Mathematical Sciences, Royal Institution, W1S 4BS, UK}
\affiliation{Center of Mathematical Sciences and Applications, Harvard University, MA 02138, USA}


\begin{abstract} 

In the standard lore, the baryon asymmetry of the present universe is attributed to the leptogenesis from the sterile right-handed neutrino with heavy Majorana fermion mass decaying into the Standard Model's leptons at the very early universe 
--- called the Majorana leptogenesis;
while the electroweak sphaleron causes 
 baryogenesis at a later time. In this work, we propose a new mechanism, named topological leptogenesis, to explain the lepton asymmetry.
Topological leptogenesis replaces 
some of the sterile neutrinos
by 
introducing a new gapped topological order sector (whose low-energy exhibits topological quantum field theory
with long-range entanglement) 
that can cancel the baryon minus lepton 
$({\bf B} - {\bf L})$
mixed gauge-gravitational anomaly of the Standard Model.
Then the Beyond-the-Standard-Model
dark matter consists of topological quantum matter,
such that the gapped non-particle excitations of
extended line and surface defect 
with fractionalization and anyon charges
can decay into the Standard Model particles.
In addition, gravitational leptogenesis can be regarded as an intermediate step (between 
Majorana particle leptogenesis 
and topological non-particle leptogenesis)
to mediate such decaying processes from the 
highly entangled gapped topological order excitations.

\end{abstract}


\maketitle

\tableofcontents



\section{Introduction and Summary}

Leptogenesis is a hypothetical physical process that produces lepton and antilepton number asymmetry 
in the very early universe.
Baryogenesis is yet another hypothetical process during the early universe to 
produce baryon and antibaryon asymmetry.
The two syntheses result in the present-day dominance of leptons over antileptons 
and baryon over antibaryons in the observed universe.
Sakharov \cite{Sakharov:1967dj} proposed three necessary conditions to produce 
an imbalance of baryons and antibaryons:
(1) Baryon number $\bB$ violation -- namely the continuous $\U(1)_{\bB}$ breaking 
(2) Discrete charge conjugation C symmetry and CP (or time-reversal T) 
symmetry violation. (3) These symmetry violations happen when the universe is 
out of thermal equilibrium.
The standard Standard Model (SM) does not contribute enough to the condition (2) and (3), 
thus the leptogenesis requires some beyond the Standard Model (BSM) effect.

One of the most popular scenarios is the {\bf Majorana leptogenesis} advocated by Fukugita and Yanagida \cite{Fukugita:1986hr}, 
among others (references therein \cite{Davidson:2008bu0802.2962}). 
 Majorana leptogenesis \cite{Fukugita:1986hr}
introduces some hypothetical heavy sterile right-handed neutrino $\nu_R$
that are sterile to the Standard Model (SM) gauge force,
while $\nu_R$ is paired with itself with Majorana mass term in the lagrangian form
$M {\nu_R}^{\rm T} {\nu_R} + {\rm h.c.}$ 
that breaks the fermion number conservation $\U(1)_{\rm F}$ down to the fermion parity $\Z_2^{\rm F}$ subgroup.
The decay of heavy $\nu_R$ can convert to 
Higgs boson and other experimentally observed less-heavy leptons ---
including 
$\U(1)_{\rm EM}$ charged leptons such as electron $e^-$, muon $\mu^-$ and tauon $\tau^-$, 
and $\U(1)_{\rm EM}$
neutral left-handed neutrinos $\nu_{L,e}$, $\nu_{L,\mu}$ and $\nu_{L,\tau}$.
Some of these leptons can further decay to quarks through 
the electroweak sphaleron 
\cite{PhysRevD.30.2212KlinkhamerMantonSphaleron, Kuzmin:1985mm} via the Adler-Bell-Jackiw (ABJ) anomaly
\cite{Adler1969gkABJ, Bell1969tsABJ} under the Yang-Mills 
field instanton \cite{tHooft1976ripPRL, BelavinBPST1975} 
of the weak gauge force, which is a popular scenario of {\bf baryogenesis}.

Another scenario of leptogenesis without necessarily introducing any sterile right-handed neutrino $\nu_R$ is the {\bf gravitational leptogenesis} \cite{Alexander:2004usPeskin0403069, Adshead1711.04800:2017znw}. The curved spacetime ripple 
locally may give rise to a source of gravitational instanton globally.
So the mixed gravitational anomaly 
\cite{Eguchi:1976db, AlvarezGaume1983igWitten1984, Putrov:2023jqi2302.14862}
between the lepton number $\U(1)_{\bf L}$
and the gravitational source from the triangle Feynman diagram 
$\U(1)_{\bf L}$-gravity$^2$ can generate an unbalanced new lepton number 
out of the curved spacetime ripple. Namely, the lepton number nonconservation or violation
can be attributed to the mixed gravitational anomaly.

The purpose of this work is to propose a new leptogenesis mechanism --- {\bf topological leptogenesis} ---  
by introducing 
a hidden BSM topological sector  that couples to the SM 
via topological gauge interaction \cite{JW2006.16996, JW2008.06499, JW2012.15860, Cheng:2024awi2411.05786}.
Precisely, the BSM topological sector
is a Topological Quantum Matter sector
whose low energy has a Topological Quantum Field Theory (TQFT, in the sense of 
Schwarz-type  
Chern-Simons-Witten-like theory 
\cite{ChernSimons1974ft, Schwarz1978cn, Witten1988hfJonesQFT}) and whose zero temperature phase (in terms of quantum many-body condensed matter sense) is {\bf 
topological order} \cite{Wen2016ddy1610.03911} (here sometimes abbreviated as TO). 
We propose that the fractionalized anyon like excitations
\cite{Wilczek:1982wyPRLAnyon, Wilczek1990BookFractionalstatisticsanyonsuperconductivity} with an energy $E$ 
above the topological order gap $\Delta_{\rm TO}$ 
(so $E \geq \Delta_{\rm TO}$)
can decay into the SM's ``elementary'' particles
via topological discrete gauge interaction (later in \eq{eq:Topo-discrete-gauge-LSM})
or via the mixed baryon minus lepton $({\bf B} - {\bf L})$
gauge-gravitational anomaly process (later in \eq{eq:TopogravgravLSM}).

Regardless of which types of leptogenesis out of the three (Majorana, gravitational, or topological)  scenarios that we decide to study,
the three scenarios can be motivated by the same common theme
on the quantum anomaly structure of the SM.
In other words, subconsciously or inadvertently, 
all three leptogenesis scenarios try
to either implement or resolve
the following anomaly facts of the SM:
\begin{enumerate}
\item The continuous baryon {\bf B} minus lepton {\bf L} number symmetry, $\U(1)_{{\bf B}- {\bf L}}$ is preserved within the SM.
More precisely, to have a properly quantized charge, it is better to 
scale $\U(1)_{{\bf B}- {\bf L}}$ 
by a factor of color number $N_c$ as
the quark {\bf Q} number minus $N_c$ lepton {\bf L} number symmetry,
$\U(1)_{{\bf Q}- N_c {\bf L}}$, where the color number is $N_c=3$ in the SM. 
So the properly quantized charge is indeed
$q_{{\bf Q}- N_c {\bf L}} = N_c q_{{\bf B}- {\bf L}}$.
{Namely,
when we mention {$\U(1)_{{\bf B}- {\bf L}}$ symmetry and anomaly}, we really mean
$\U(1)_{{\bf Q}- N_c {\bf L}}$ symmetry and anomaly.}
But the $\U(1)_{{\bf Q}- N_c {\bf L}}$ symmetry has a 
't Hooft anomaly in 4d spacetime captured by two 
triangle Feynman diagrams:
\bea \label{eq:BL-diagram}
\includegraphics[scale=0.65]{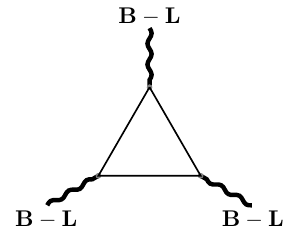}
\includegraphics[scale=0.65]{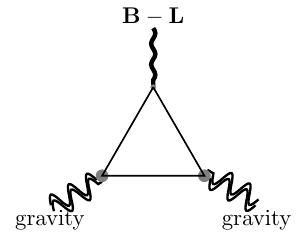}
\eea
The 4d 't Hooft anomaly is also captured by
a 5d invertible topological field theory (iTFT) in one extra dimension
with the following invertible complex U(1) valued functional via the anomaly inflow \cite{1984saCallanHarvey, WangWanYou2112.14765, WangWanYou2204.08393, Putrov:2023jqi2302.14862}:
\begin{equation}\label{SM-U1-iTFT-1}
   \exp(\ii  S_5)
\equiv \exp\Bigg[\ii   \int_{M^5} (-N_f+n_{\nu_R}) \,A \wedge \left(N_c^3 \frac{1}{6} \dd A \wedge  \dd A +N_c\frac{1}{24}  \frac{1}{8 \pi^2}   \Tr[ R \wedge R] \right) \Bigg], 
\end{equation} 
where the anomaly index 
$-N_f+n_{\nu_{R}}$ counting the difference between the family 
$N_f =3$ and the total number of the types of right-hand neutrinos 
$n_{\nu_{R}} \equiv \sum_{\rm I} n_{\nu_{{\rm I},R}}$.  
The $n_{\nu_{R}}$ is so far undetermined by the experiments,
so $n_{\nu_{R}}$
can be equal, smaller, or larger than 3 (here ${\rm I}=1,2,3,\dots$ for $e,\mu,\tau,\dots$ type of neutrinos). 
The $A$ is the $\U(1)_{{\bf Q}- N_c {\bf L}}$ 
gauge field connection (locally a 1-form),
and $R$ is the spacetime curvature locally a 2-form.
The $\U(1)^3$ anomaly in 4d matches with
the 5d $A \wedge \dd A \wedge  \dd A$ term,
and the $\U(1)$-gravity$^2$ anomaly in 4d
matches with the 5d $A \wedge  \Tr[ R \wedge R]$ term.

We shall beware that the 
 two Feynman diagrams from
$\U(1)_{\bf B}^3$ and 
$\U(1)_{\bf B}$-gravity$^2$ contribute zero anomaly indices, schematically:
$$ \label{eq:B-diagram}
\includegraphics[scale=0.65]{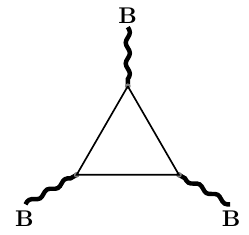} 
\includegraphics[scale=0.65]{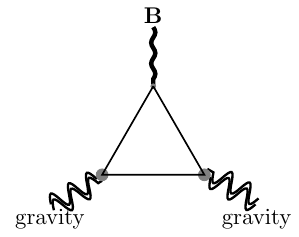} 
$$
So the only contribution of \eq{eq:BL-diagram} and \eq{SM-U1-iTFT-1}
is from $\U(1)_{\bf L}^3$ and 
$\U(1)_{\bf L}$-gravity$^2$ anomalies:
\bea \label{eq:L-diagram}
\includegraphics[scale=0.65]{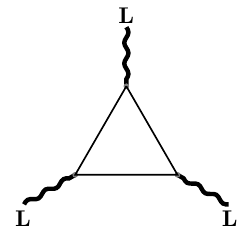}
\includegraphics[scale=0.65]{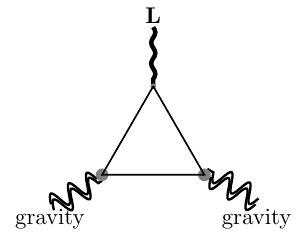}
\eea
The 4d anomaly of \eq{eq:L-diagram} is captured by
the 5d iTFT similar to \eq{SM-U1-iTFT-1} but a rescaling with
$A_{\bf L} = - N_c A$.
The reason 
we prefer to focus on 
$\U(1)_{{\bf Q}- N_c {\bf L}}$ instead of $\U(1)_{\bf L}$
symmetry is because only the former $\U(1)_{{\bf Q}- N_c {\bf L}}$ has
mixed anomaly-free with the SM gauge group 
but the latter $\U(1)_{\bf L}$ is anomalous with the SM gauge group.
This relates to the next anomaly fact.

\item  $\U(1)_{{\bf B}- {\bf L}}$ (or more precisely
$\U(1)_{{\bf Q}- N_c {\bf L}}$)
is mixed anomaly-free with the SM Lie algebra 
$\cG_{\rm SM} \equiv su(3) \times su(2) \times u(1)_{\tilde Y}$,
or any of the four versions of the SM gauge group 
$G_{\rm SM} \equiv (\SU(3) \times \SU(2)  \times \U(1)_{\tilde Y})/\Z_q$
gauge group for $q=1,2,3,6$ \cite{Tong2017oea1705.01853}.
But the
$\U(1)_{\bf B}$ or $\U(1)_{\bf L}$ alone, each individually,
does have mixed anomaly with $su(2) \times u(1)_{\tilde Y}$.
We write 
$N_f=3$ families of 15 or 16 Weyl fermions (spin-$\frac{1}{2}$ Weyl spinor 
is in the ${\bf 2}_L^\C$ representation {of} the spacetime symmetry Spin(1,3)
as a left-handed 15- or 16-plet $\psi_L$)
in the following $\cG_{\rm SM}$ representation
\begin{multline}
    \label{eq:SMrep}
({\psi_L})_{\rm I} =
( \bar{d}_R \oplus {l}_L  \oplus q_L  \oplus \bar{u}_R \oplus   \bar{e}_R  
)_{\rm I}
\oplus
n_{\nu_{{\rm I},R}} {\bar{\nu}_{{\rm I},R}}
\\
\sim 
\big((\overline{\bf 3},{\bf 1})_{2} \oplus ({\bf 1},{\bf 2})_{-3}  
\oplus
({\bf 3},{\bf 2})_{1} \oplus (\overline{\bf 3},{\bf 1})_{-4} \oplus ({\bf 1},{\bf 1})_{6} \big)_{\rm I}
\oplus n_{\nu_{{\rm I},R}} {({\bf 1},{\bf 1})_{0}}
\end{multline} 
for each family. 
The family index is in the roman font ${\rm I}=1,2,3$; 
with ${\psi_L}_1$ for $u,d,e$ type,
${\psi_L}_2$ for $c,s,\mu$ type,
and 
${\psi_L}_3$ for $t,b,\tau$ type of quarks and leptons.
Thus the following Feynman diagram consists of two
non-vanishing ABJ anomaly contributions
\bea \label{eq:LgravgravB}
\includegraphics[scale=0.8]{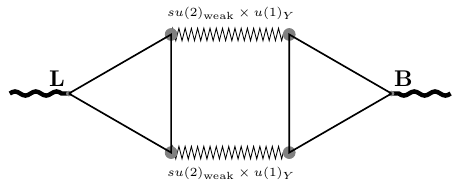},
\eea
which can convert 
between $\U(1)_{\bf B}$ and $\U(1)_{\bf L}$ number violation
through their anomalous current formulae:
\bea \label{eq:dJQL}
\dd \star j_{\bf Q} &=& - N_c N_f (18 \frac{ c_1(\U(1)_{\tilde{Y}})^2}{2} +c_2(\SU(2))). \cr
\dd \star j_{\bf L} &=&  - N_f (18 \frac{ c_1(\U(1)_{\tilde{Y}})^2}{2} +c_2(\SU(2))). 
\eea
The anomalous current requires either nontrivial
instanton number from the second Chern class $c_2$ globally 
of $su(2)$ (the sphaleron)
 or the first Chern class $c_1^2$ of $u(1)$:
This \eq{eq:LgravgravB} 
schematically illustrates the {\bf electroweak sphaleron baryogenesis} conversion process between ${\bf B}$ and ${\bf L}$.\footnote{The $c_1$ is obtained 
by integrating over 
$c_1 \equiv \int \frac{1}{2 \pi} \Tr {F}$ locally
with 
nontrivial transition functions between patches.
The $c_2$ is obtained 
by integrating over 
$c_2 \equiv \int -\frac{1}{8\pi^2} \Tr({F}\wedge {F}) 
+\frac{1}{2} (\frac{1}{2 \pi} \Tr {F})^2$ locally with 
nontrivial transition functions between patches.
It is more common to consider the $su(2)$ sphaleron because
the instanton number on a compact space like 4-sphere $S^4$ can still be non-zero. 
But the $u(1)$ instanton number on the $S^4$ 
is trivial, only that on those such as 
$S^2 \times S^2$ with nontrivial lower homotopy
can be nonzero.}

\end{enumerate}

In the following sections, 
by using the above two anomaly facts,
we can streamline the known
scenarios of
{\bf gravitational leptogenesis} 
in \Sec{sec:gravitationalleptogenesis}
and 
{\bf Majorana leptogenesis} in \Sec{sec:Majoranaleptogenesis},
then we propose the {\bf topological leptogenesis} in \Sec{sec:Topologicalleptogenesis}.
Table \ref{table:leptogenesis} summarizes, 
compares, and contrasts the three scenarios.

\section{Gravitational Leptogenesis} 
\label{sec:gravitationalleptogenesis}

Gravitational leptogenesis \cite{Alexander:2004usPeskin0403069,Adshead1711.04800:2017znw}
can be simply explained by the following Feynman diagram
\bea \label{eq:gravgravLSM-diagram}
\includegraphics[scale=0.8]{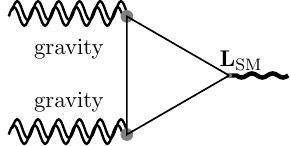}.
\eea
In the pure gravitational leptogenesis,
we have 
$$N_f=3 \text{ and } n_{\nu_R}=0.$$
The anomalous current nonconservation 
(or the violation of the $\U(1)$ number conservation for 
the lepton number ${\bf L}$ or ${{\bf Q}- N_c {\bf L}}$ of the SM)
obeys:
\bea \label{eq:dJQL-p1-RR}
\dd \star j_{{\bf L}} &=& 
(-N_f+n_{\nu_R}) \frac{p_1(TM)}{24}.\cr
\dd \star j_{{\bf Q}-N_c {\bf L}} &=& 
-(-N_f+n_{\nu_R}) \,N_c\frac{p_1(TM)}{24}.
\eea
The lepton number ${\bf L}$ here in \eq{eq:dJQL-p1-RR} 
is the ${\bf L}_{\SM}$ of SM in \eq{eq:gravgravLSM-diagram}.  
These equations can also be solved from the variation of the bulk-boundary coupled equation of \eq{SM-U1-iTFT-1}.
The first Pontryagin class $p_1$ can be defined globally
when integrating over $\Tr[ R \wedge R]$ locally over the spacetime 4-manifold, by taking into account the transition functions between overlapping patches. The $p_1$ is also related to the gravitational Chern-Simons 3-form:
\bea
\label{eq:Pontryagin}
p_1 &\coloneqq&- \frac{1}{8 \pi^2}   \Tr[ R \wedge R]
= - \frac{1}{8 \pi^2}   R^a{}_b \wedge R^b{}_a
= - \frac{1}{8 \pi^2}
\frac{1}{2^2}
\tilde\epsilon^{\mu \nu \al \bt} R^a{}_b{}_{\mu \nu}  R^b{}_a{} {}_{\al \bt} \dd^4 x
\equiv -\frac{1}{2\pi} \dd \GCS.\\
\label{eq:GCS}
\GCS &\coloneqq&  \frac{1}{4\pi}\Tr[\omega\wedge d\omega+
    \frac{2}{3}\,\omega\wedge \omega\wedge \omega]
=\frac{1}{4\pi}
\tilde\epsilon^{ \nu \al \bt} 
(\omega_{\nu}{}^a{}_b \prt_\al \omega_{\bt}{}^b{}_a 
+ 
\frac{2}{3}\, \omega_{\nu}{}^a{}_b \omega_{\al }{}^b{}_c \omega_{\bt}{}^c{}_a
) \dd^3 x.
\eea
In \eq{eq:dJQL-p1-RR}, 
the local nonzero $\Tr[ R \wedge R]$ 
can make $\dd \star j_{{\bf L}}$ nonzero locally.
However, to achieve the lepton number violation globally in the early universe,
we need to have the spacetime topology changing process 
so that the gravitational instanton number $p_1$
changes (such as creating
a K3 surface with $p_1({\rm K3})=-16$ or $p_1(\CP^2)=1$ in the Euclidean signature). So the lepton number changing from the time $t_i$ to $t_f$
via a topology changing $\Delta M^4$ 
is 
\bea
L (t_f) -  L (t_i) \equiv 
\Delta L = \int_{\Delta M^4}\dd \star j_{{\bf L}} = 
(-N_f+n_{\nu_R}) \frac{\Delta p_1}{24}.
\eea
There are a few challenges for this scenario:
\begin{enumerate}
\item
Gravitational leptogenesis by itself \cite{Alexander:2004usPeskin0403069}
originally does not provide light neutrino mass.
The light neutrino mass here requires a new mechanism.
Proposals along this direction can be found in 
\cite{Dvali:2016uhn1602.03191, Dvali:2021uvk2112.02107}.

\item The continuous U(1) of  ${({\bf B} -  {\bf L})}$ is  
 unbroken from the IR to the deep UV,
 thus it has to be dynamically gauged to be consistent with the quantum gravity argument (References therein \cite{McNamara2019rupVafa1909.10355}). 
But a new U(1) photon of dynamically gauged
gauge ${({\bf B} -  {\bf L})}$ contradicts with experiments in nature.

\item  The model itself is gravitational anomalous, so this proposal alone
may not be a final story in a dynamical gravity theory 
or in full quantum gravity. The full theory better to be fully anomaly-free 
thus in a trivial cobordism class \cite{McNamara2019rupVafa1909.10355}
in the sense of  \cite{WangWanYou2112.14765, WangWanYou2204.08393}.
\end{enumerate}
Majorana leptogenesis can rescue some of these challenges.

\section{Majorana Leptogenesis}
\label{sec:Majoranaleptogenesis}

The popular  Majorana leptogenesis 
\cite{Fukugita:1986hr, Davidson:2008bu0802.2962}
provides a UV completion of gravitational leptogenesis
by canceling the gravitational anomaly via adding $\nu_R$.
For a pure Majorana leptogenesis to cancel the gravitational anomaly, 
we have 
$$N_f=3 \text{ and } n_{\nu_R}=3.$$
The gravitational anomaly was not emphasized or motivated in the original
Fukugita-Yanagida model \cite{Fukugita:1986hr}, 
thus it is common that the traditional 
leptogenesis does not demand $n_{\nu_R}=3$
\cite{Davidson:2008bu0802.2962}.
A schematic Feynman diagram that evolves Majorana leptogenesis through
gravitational anomaly can be: 
\bea
\includegraphics[scale=0.8]{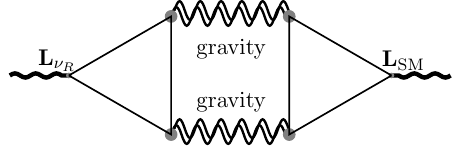}
\eea
But this gravitational involved effect can be a much less significant order 
effect namely a higher-order effect.
Indeed
the heavy $\nu_R$ and its lepton number current of
${\bf L}_{\nu_R}$
can decay to 
Higgs boson and other experimentally observed less-heavy leptons ---
including 
neutral left-handed neutrinos $\nu_{L,e}$, $\nu_{L,\mu}$ and $\nu_{L,\tau}$,
and the Higgs can further decay to other quarks and leptons
via the following terms \eq{eq:YHnu} and \eq{eq:nuRmass} 
\cite{Fukugita:1986hr}.
A typical seesaw mechanism 
introduces 
Dirac mass term with the SM Higgs $\phi_H$
and with Yukawa-Higgs coupling $\lambda_\nu$,
\bea \label{eq:YHnu}
\lambda_\nu \bar\nu_R  \phi_H^\dagger  \nu_L + \lambda_\nu^* \bar\nu_L  \phi_H  \nu_R,
\eea
and
Majorana mass term among $\nu_R$,
\bea \label{eq:nuRmass}
\frac{\ii M }{2}( \nu_R^{\rm T} \sigma^2 \nu_R - \nu_R^\dagger  \sigma^2  \nu_R^*  ).
\eea
\Eq{eq:nuRmass} explicitly 
breaks the neutrino number 
continuous symmetry of $\U(1)$ down to the discrete fermion parity $\Z_2^\rF$.
If the $\nu_R$ is heavy with a large Majorana mass $M$,
the nearly massless energy eigenstate has 
$\frac{m^2}{M} \sim
\frac{| \lambda_\nu \< \phi_H\>   |^2}{M}$.
The low-energy physics by integrating out the heavy $\nu_R$ gives rise to
the dimension-5 Weinberg operator
\bea \label{eq:nuLmass}
\sim \frac{1}{M} (\lambda_\nu^2 \cdot
\nu_L^{\rm T} \phi_H^*  \phi_H^\dagger  \nu_L + 
\lambda_\nu^{*2}\cdot
\nu_L^{\dagger} \phi_H  \phi_H^{\rm T}  \nu_L^*)
\eea
that gives the effective Majorana mass to $\nu_L$.

In Majorana leptogenesis 
\cite{Fukugita:1986hr, Davidson:2008bu0802.2962},
the $\nu_R$ replaces 
the role of the curved spacetime ripple with nonzero $\Tr[ R \wedge R]$
in gravitational leptogenesis. But there are some curious facts about 
Majorana leptogenesis that may pose challenges to it:
\begin{enumerate}
\item The $\nu_R$ Majorana mass term already breaks the U(1) version of
${\bf B} -  {\bf L}$ (namely $\U(1)_{{\bf Q}- N_c {\bf L}}$)
down to fermion parity $\Z_2^\rF$ at its mass energy scale $M$ at UV.
So it is curious how that the broken ${\bf B} -  {\bf L}$ 
\emph{re-emerge}
at the SM scale as a good nearly-anomaly-free continuous global symmetry 
$\U(1)_{{\bf Q}- N_c {\bf L}}$ back in the SM at IR.

From the renormalization group (RG) and IR effective field theory (EFT) viewpoint in \eq{eq:nuLmass}, we may say that the
$\U(1)_{{\bf Q}- N_c {\bf L}}$ is never an exact global symmetry
but only an approximate symmetry, broken by the ratio of $O(\frac{E}{M})$ at the energy scale $E$.

\item Once the U(1) version of
${\bf B} -  {\bf L}$ (namely $\U(1)_{{\bf Q}- N_c {\bf L}}$)
is broken by $\nu_R$ Majorana mass term at UV,
there imposes no direct IR constraint to demand the 16th Weyl fermion
$\nu_R$ to cancel the SM's ${\bf B} -  {\bf L}$ anomaly.
Because there is no ${\bf B} -  {\bf L}$ anomaly at UV when ${\bf B} -  {\bf L}$ is already broken at UV.

So it seems that more nature to have an alternative 
$({\bf B} -  {\bf L})$-preserving scenario at UV 
with an alternative energetically gapped state, but without 
introducing $\nu_R$.

\end{enumerate}

These curious facts prompt the author to think of an alternative scenario:
topological leptogenesis. 
The author pursues the alternative story such that 
the discrete version of ${\bf B} -  {\bf L}$ 
can be exact, anomaly-free, and gaugeable in \Sec{sec:Topologicalleptogenesis}.
\Table{table:mass-symmetry-preserving-breaking} summarizes the
symmetry-preserving vs symmetry-breaking patterns for these scenarios.

\begin{table}[!h]
\begin{tabular}{|c|c|c|c| c |}
    \hline
& {\parbox{3.5cm}{ 
\vspace{2pt}
\centering  neutrino Yukawa-Higgs\\ Dirac mass\\
\eqq{eq:YHnu}
\vspace{2pt}}}  
& 
 {\parbox{3.5cm}{ 
\vspace{2pt}
$\nu_R$ Majorana mass\\ 
\eqq{eq:nuRmass} 
\vspace{2pt}}}& 
 {\parbox{3.5cm}{ 
\vspace{2pt}
\centering  $\nu_L$ dimension-5 \\ 
Majorana mass \\
\eqq{eq:nuLmass}
\vspace{2pt}}}
& 
 {\parbox{3.5cm}{ 
\vspace{2pt}
$\Z_{4,X}$-symmetric \\
TO / TQFT \\
or CFT \cite{JW2006.16996, JW2008.06499, JW2012.15860, Cheng:2024awi2411.05786}
\vspace{2pt}}}
\\ 
    \hline
    \hline
 {\parbox{1.8cm}{ 
\vspace{2pt}
Continuous\\
$\U(1)_{\tilde Y}$ 
\vspace{2pt}}}
& 
 {\parbox{3.6cm}{ 
 \raggedright  
\vspace{2pt}
Explicit: preserved.\\
Spontaneous: broken\\
by $\<\phi_H\>$ SSB whose ${\tilde Y}=3$
below $E \lesssim  \Lambda_{\rm EW}$.
\vspace{2pt}}}
&
{\parbox{3.6cm}{ 
 \raggedright  
\vspace{2pt}
Explicit: preserved.
\vspace{2pt}}}
& 
{\parbox{3.6cm}{ 
 \raggedright  
\vspace{2pt}
Explicit: preserved.\\
Spontaneous: broken\\
by $\<\phi_H\>$ SSB below $E \lesssim  \Lambda_{\rm EW}$.
\vspace{2pt}}}
& {\parbox{3.6cm}{ 
 \raggedright  
\vspace{2pt}
Explicit: preserved.\\
Spontaneous: preserved.
\vspace{2pt}}} \\
\hline
 {\parbox{1.8cm}{ 
\vspace{2pt}
Continuous\\
${({\bf B} -  {\bf L})}$ or\\
$\U(1)_{{\bf Q}- N_c {\bf L}}$
\vspace{2pt}}}
& 
{\parbox{3.6cm}{ 
 \raggedright  
\vspace{2pt}
Explicit: preserved.\\
Spontaneous: preserved.
\vspace{2pt}}}
& 
{\parbox{3.6cm}{ 
 \raggedright  
\vspace{2pt}
Explicit: broken,\\
down to $\Z_2^\rF$.
\vspace{2pt}}}
& 
{\parbox{3.6cm}{ 
 \raggedright  
\vspace{2pt}
Explicit: broken,\\
down to $\Z_2^\rF$.
\vspace{2pt}}}
& 
{\parbox{3.6cm}{ 
 \raggedright  
\vspace{2pt}
Explicit: No continuous ${({\bf B} -  {\bf L})}$ symmetry.\\
\vspace{2pt}}}
\\
\hline
 {\parbox{1.8cm}{ 
\vspace{2pt}
Continuous\\
$\U(1)_{X}$ 
\vspace{2pt}}}
& 
{\parbox{3.6cm}{ 
 \raggedright  
\vspace{2pt}
Explicit: preserved.\\
Spontaneous: broken\\
by $\<\phi_H\>$ SSB whose $X=-2$
below $E \lesssim  \Lambda_{\rm EW}$ down to $\Z_2^\rF$.
\vspace{2pt}}}
& 
{\parbox{3.6cm}{ 
 \raggedright  
\vspace{2pt}
Explicit: broken,\\
down to $\Z_2^\rF$.
\vspace{2pt}}}
& 
{\parbox{3.6cm}{ 
 \raggedright  
\vspace{2pt}
Explicit: broken,\\
down to $\Z_2^\rF$.\\
Spontaneous: broken already, 
but also by $\<\phi_H\>$ SSB whose $X=-2$
below $E \lesssim  \Lambda_{\rm EW}$.
\vspace{2pt}}}
& 
{\parbox{3.6cm}{ 
 \raggedright  
\vspace{2pt}
Explicit: No continuous $\U(1)_{X}$  symmetry.\\
\vspace{2pt}}}
\\
\hline
 {\parbox{1.8cm}{ 
\vspace{2pt}
Discrete\\
$\Z_{4,X}$ 
\vspace{2pt}}}
& 
\parbox{3.6cm}{ 
 \raggedright  
\vspace{2pt}
Explicit: preserved.\\
Spontaneous: broken\\
by $\<\phi_H\>$ SSB whose $X=2$
below $E \lesssim  \Lambda_{\rm EW}$ down to $\Z_2^\rF$.
\vspace{2pt}}
& 
{\parbox{3.6cm}{ 
 \raggedright  
\vspace{2pt}
Explicit: broken,\\
down to $\Z_2^\rF$
\vspace{2pt}}}
& 
{\parbox{3.6cm}{ 
 \raggedright  
\vspace{2pt}
Explicit: broken,\\
down to $\Z_2^\rF$.\\
Spontaneous: broken already, 
but also by $\<\phi_H\>$ SSB whose $X=2$
below $E \lesssim  \Lambda_{\rm EW}$.
\vspace{2pt}}}
& 
{\parbox{3.6cm}{ 
 \raggedright  
\vspace{2pt}
Explicit: preserved.\\
Discrete
$\Z_{4,X}$ is exact, anomaly-free gaugeable for the full SM + BSM
\cite{JW2012.15860}, and dynamically gauged in a full quantum gravity theory
\cite{McNamara2019rupVafa1909.10355, WangWanYou2112.14765, WangWanYou2204.08393}.
 \vspace{2pt}}}
\\
\hline
\end{tabular}
\caption{Comparison of four different mass generating physics,
and their symmetry-preserving or symmetry-breaking patterns.
SSB stands for the conventional ``spontaneous symmetry breaking.'' 
The comment on the ``Spontaneous'' only applies to those terms
with $\<\phi_H\>$ SSB for the energy scale below the electroweak scale  
$E \lesssim  \Lambda_{\rm EW}$.
}
\label{table:mass-symmetry-preserving-breaking}
\end{table}  

\newpage

\section{Topological Leptogenesis}
\label{sec:Topologicalleptogenesis}

Indeed when the ${\bf B} -  {\bf L}$ is discrete, such as
a discrete $\Z_{4,X}$ subgroup, 
where $X \equiv 
5({ \mathbf{B}-  \mathbf{L}})-\frac{2}{3} {\tilde Y}
=\frac{5}{N_c}({ \mathbf{Q}- N_c  \mathbf{L}})-\frac{2}{3} {\tilde Y}$ 
with properly integer quantized hypercharge $\tilde Y$ \cite{Wilczek1979hcZee, WilczekZeePLB1979}, 
the perturbative mixed ${\bf B} -  {\bf L}$-gravitational anomaly
with a $\Z$ classification
becomes a nonperturbative mixed $\Z_{4,X}$-gravitational anomaly
with a $\Z_{16}$ classification for this anomaly in 4d spacetime
\cite{2018arXiv180502772T, GarciaEtxebarriaMontero2018ajm1808.00009, Hsieh2018ifc1808.02881, GuoJW1812.11959, WW2019fxh1910.14668}.
There are new ways to cancel the $\Z_{16}$ anomaly by introducing
4d TQFT or 4d CFT and 5d invertible TQFT \cite{JW2006.16996, JW2008.06499, JW2012.15860, Cheng:2024awi2411.05786}.
Importantly, these new 4d TQFT or 4d CFT and 5d invertible TQFT preserve the $\Z_{4,X}$ entirely without $\Z_{4,X}$-breaking.
The underlying mechanism of these new interacting sectors is based on the symmetry-extension \cite{Wang2017locWWW1705.06728}, 
instead of the symmetry-breaking of
the traditional Nambu-Goldston-Anderson-Higgs or Landau-Ginzburg symmetry-breaking type theory.
So topological leptogenesis introduces some new potential $\Z_{4,X}$-symmetric
gapped topological order with low energy 4d TQFT.
The characteristic properties of {\bf 
topological order} are:
\begin{enumerate}
\item
The energy spectrum above the TQFT ground state 
 is gapped, with an energy gap size denoted 
 $\Delta_{\rm TO}$. 
The energetic excitations with energy $E$ above the gap
 ($E \geq \Delta_{\rm TO}$)
contain fractionalized excitations with anyon statistics \cite{Wilczek1990BookFractionalstatisticsanyonsuperconductivity}
(that can be neither boson nor fermion statistics)

\item The ground state degeneracy (the number of ground states) of 
TO / TQFT depends on the spatial topology. The ground states of TO / TQFT
exhibit long-range entanglement that cannot be deformed to a trivial gapped vacuum (with a single ground state) via any local quantum unitary transformation.
\end{enumerate}
We shall name this scenario topological leptogenesis, due to its 
long-range entanglement property of TO / TQFT. Topological leptogenesis
resolves:
\begin{enumerate}
\item A UV completion of gravitational leptogenesis without $\nu_R$ as this schematic process:
\bea \label{eq:TopogravgravLSM}
\includegraphics[scale=0.8]{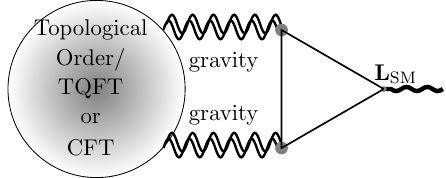}.
\eea
\item Furthermore, without using gravity as an intermediate mediator,
there is also the discrete ${\bf B} -  {\bf L}$ gauge topological force (the
$\Z_{4,X}$ gauge force) to mediate between the $\Z_{4,X}$-charged 
quarks and leptons in the SM and the $\Z_{4,X}$-Cheshire-charged extended
TO / TQFT link configurations (that stores the $\Z_{4,X}$ charge nonlocally
in the link in the Cheshire-charge way):
\bea \label{eq:Topo-discrete-gauge-LSM}
\includegraphics[scale=0.8]{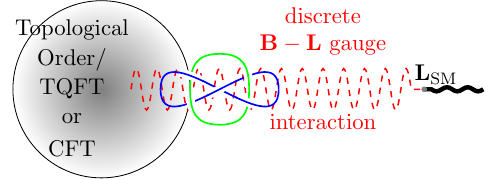}.
\eea
Follow \cite{GuoJW1812.11959}, here we show the Whitehead link that can detect a similar kind of
Arf invariants \cite{Arf1941}
and Arf-Brown-Kerverie invariant \cite{Brown-Kervaire, Kapustin1406.7329} appear in the 4d TQFT construction \cite{JW2006.16996, JW2008.06499, JW2012.15860, Cheng:2024awi2411.05786}.

\item
Different low-energy physics at IR:
This scenario cannot generate the dimension-5 Weinberg operator, 
because that operator breaks the continuous and discrete $X$ symmetry ($\U(1)_X$ and $\Z_{4,X}$) explicitly. 
Also the TQFT / TO have fundamentally different properties than the defects studied in \cite{Dvali:2016uhn1602.03191, Dvali:2021uvk2112.02107}.

\end{enumerate}

\onecolumngrid

\begin{table}[!h]
\hspace{0.mm}
    \setlength{\extrarowheight}{2pt}
 \begin{tabular}{|c|c|c|c|}
    \hline
  &  Majorana leptogenesis \cite{Fukugita:1986hr} & Gravitational leptogenesis 
    \cite{Alexander:2004usPeskin0403069} & Topological leptogenesis \cite{JW2012.15860} \\
     \hline
     \hline
$\nu_R$ & Require $\nu_R$ & No $\nu_R$ needed & No $\nu_R$ needed  \\
     \hline
$\begin{array}{c}
\text{UV symmetry}\\ 
\text{\footnotesize ($E$ around $M$}\\
\text{\footnotesize  or $\Delta_{\rm TO}$)}
\end{array}$
& 
$\begin{array}{c}
\text{${({\bf B} -  {\bf L})}$ 
and $\U(1)_X$ alike,}\\
\text{and $\Z_{4,X}$ \emph{explicitly broken} by}\\
\text{$\nu_R$'s Majorana mass term.}
\end{array}$
&  
$\begin{array}{c}
\text{${({\bf B} -  {\bf L})}$, $\U(1)_X$, and $\Z_{4,X}$}\\
\text{ preserved}\\
\text{(no Majorana mass term).}
\end{array}$
&  
{\parbox{6.7cm}{ 
\vspace{2pt}
        \centering  
{${({\bf B} -  {\bf L})}$, $\U(1)_X$, and $\Z_{4,X}$ preserved.}\\
A $X$-symmetric gapped topological order \\
{with a low-energy TQFT\\
(no Majorana mass term).}\\
{Discrete $X$ becomes dynamically gauged \cite{KraussWilczekPRLDiscrete1989}}\\
{due to no global symmetry\\
in quantum gravity \cite{McNamara2019rupVafa1909.10355}.}
\vspace{2pt}
}}
\\
\hline
$\begin{array}{c}
\text{IR symmetry}\\ 
\text{\footnotesize ($E$ below $\Lambda_{\rm EW}$)}\\ 
\text{ }
\end{array}$
& 
\multicolumn{3}{ c|}{\parbox{16cm}{ 
        \centering  
        Continuous and also discrete ${({\bf B} -  {\bf L})}$ are fully preserved in the SM. But the continuous and discrete $X$\\
        are \emph{spontaneously broken} down to $\Z_2^\rF$ due to Higgs condensate $\< \phi_H\>$ 
        carries $X$ and electroweak hyper $Y$ charges. \\
    }}  
\\
\hline
{\parbox{2.cm}{ 
        \centering 
Anomaly\\
matching or\\
cancellation\\[1mm]        
}}        
 & 
 {\parbox{4.1cm}{ 
 \vspace{2pt}
        \raggedright 
$\bullet$ Adding $\nu_R$ cancels the continuous ${({\bf B} -  {\bf L})}$-grav$^2$
and ${({\bf B} -  {\bf L})^3}$ anomalies, and the discrete $\Z_{4,X}$-gravity anomaly.\\
$\bullet$ But $\nu_R$'s Majorana mass \emph{already breaks}
the ${({\bf B} -  {\bf L})}$ and $X$ symmetries at UV.
\vspace{2pt}
}}  
 & 
 {\parbox{4.1cm}{ 
        \raggedright 
$\bullet$ The continuous ${({\bf B} -  {\bf L})}$-grav$^2$
and ${({\bf B} -  {\bf L})^3}$ anomalies, and the discrete $\Z_{4,X}$-gravity anomaly
are \emph{not} canceled.\\[2.5mm]
$\bullet$ Gravitational anomaly generates leptogenesis.\\ 
}}
 & 
 {\parbox{6.7cm}{ 
\vspace{2pt}
\raggedright
$\bullet$  The continuous 
U(1) of  ${({\bf B} -  {\bf L})}$ or $X$ is \emph{not} exact \emph{nor} dynamically gauged at UV.  \\[0.5mm]
$\bullet$ The discrete $X$ (say $\Z_{4,X}$) is exact and dynamically gauged at UV. \\[0.5mm]
$\bullet$ The $\Z_{16}$ class of $\Z_{4,X}$-gravity anomaly with index $-3 \mod 16$ in the SM
is canceled by the new BSM TQFT or CFT.
}}
 \\
    \hline
$\begin{array}{c}
\text{Dark Matter}\\
\text{candidate}
\end{array}$
&  heavy $\nu_R$ with Majorana mass
   & 
{\parbox{4.cm}{ 
        \centering 
   additional axion $\varphi$ like\\
   in $\varphi\Tr[R \wedge R]$   \cite{Alexander:2004usPeskin0403069}
}
}   
   & 
{\parbox{6.7cm}{ 
\vspace{2pt}
\raggedright $\bullet$ Fractionalized 0d particle (1d worldline) and
anyon 1d loop (2d worldsheet) gapped excitations 
above $\Delta_{\rm TO}$ in the 4d spacetime.\\
$\bullet$ Gapped SPTs with low-energy iTFT in the 5d spacetime.\\
$\bullet$ Gapped topological order with low-energy noninvertible TQFT in the 5d spacetime when 
$\Z_{4,X}$ becomes dynamically gauged at UV high-energy.\\
$\bullet$ Possible unparticle CFT excitations \cite{JW2012.15860}
if we use interacting CFT to cancel the $\Z_{4,X}$-gravity anomaly.
\vspace{2pt}
}
} 
\\
\hline
Baryogenesis
& 
\multicolumn{3}{ c|}{\parbox{16cm}{ 
\vspace{2pt}
        \raggedright  
        All scenarios can use the sphaleron from the electroweak $su(2)_{\rm w} \times u(1)_{Y}$ sector to generate baryon asymmetry. \\
        $\bullet$ 
        Typically through the sphaleron doing the conversion from the anomalous lepton current via ${\bf L}$-$su(2)_{\rm w}^2$ anomaly to 
        the anomalous baryon current via ${\bf B}$-$su(2)_{\rm w}^2$ anomaly in the flat spacetime. \\
        $\bullet$ It is possible to have the
        conversion from the anomalous lepton current via ${\bf L}$-$u(1)_{Y}^2$ anomaly to 
        the anomalous baryon current via ${\bf B}$-$u(1)_{Y}^2$ anomaly in the nontrivial spacetime topology.\\
        $\bullet$ ${({\bf B} -  {\bf L})}$ current is conserved and non-anomalous, because
        ${({\bf B} -  {\bf L})}$-$su(2)_{\rm w}^2$ and ${({\bf B} -  {\bf L})}$-$u(1)_{Y}^2$ are anomaly-free,
        at this energy scale range.
          \\
          \vspace{2pt}
    }}  
\\
\hline   
{\parbox{2.cm}{ 
        \centering 
Comments      
}}        
 & 
{\parbox{4.cm}{ 
        \raggedright 
{\bf Pros}: \\
Seesaw mechanism
 with tiny $\nu_L$ mass.
Then the sphaleron 
 generates the baryon 
asymmetry.\\[1mm]
{\bf Cons}: \\
$\bullet$ No $\Z_{16}$ anomaly
 cancellation at all    
at UV (due to $\Z_{4,X}$
 already broken by
Majorana mass) to  
constrain 16th Weyl
  $\nu_R$. \\[1mm]
$\bullet$ Both continuous  
 and discrete $X$  
($\U(1)_X$ and $\Z_{4,X}$)
are \emph{explicitly broken} 
 at deep UV, but 
re-emerge at IR 
then to be SSB \emph{again} 
by Higgs vev $\< \phi_H\>$. $\Rightarrow$
 More contrived 
symmetry pattern 
from UV to IR. \\[1mm]
}} 
 & 
 {\parbox{4.cm}{ 
 \vspace{2pt}
        \raggedright 
{\bf Pros}: \\
Gravitational instanton / gravitational wave
or curved spacetime background generates lepton asymmetry.\\
Then the sphaleron generates the baryon asymmetry.\\[1mm]
{\bf Cons}: \\
$\bullet$ The continuous 
U(1) of  ${({\bf B} -  {\bf L})}$ is  
 unbroken from IR to UV,
 thus it has to be dynamically gauged to be consistent with quantum gravity \cite{McNamara2019rupVafa1909.10355}. But
new gauge U(1) photon of  ${({\bf B} -  {\bf L})}$ contradicts with experiments in nature.\\
$\bullet$  The model itself is
 gravitational 
anomalous.\\[1mm]
}} 
 & 
{\parbox{6.7cm}{ 
 \vspace{2pt}
\raggedright 
{\bf Pros}: \\
$\bullet$ 
The U(1) of ${({\bf B} -  {\bf L})}$ is \emph{not} exact \emph{nor} dynamically gauged.
No additional new gauge U(1) of ${({\bf B} -  {\bf L})}$, thus no new photon is needed.
 \\[2mm]
 $\bullet$ Consistent discrete $\Z_{4,X}$ gauge symmetry preserving from deep UV to IR,
until below $\Lambda_{\rm EW}$ scale, the $\Z_{4,X}$ SSB by $\< \phi_H\>$ condensate.\\[2mm]
$\bullet$ Consistent discrete ${({\bf B} -  {\bf L})}$ gauge symmetry preserving from deep UV to IR,
even below $\Lambda_{\rm EW}$ scale, it is still preserved.\\[2mm]
$\bullet$ Both discrete $X$ and discrete ${({\bf B} -  {\bf L})}$ are secretly dynamically gauged, although may behave as global symmetries at IR \cite{KraussWilczekPRLDiscrete1989}. The model with SM $+$ topological matter is anomaly-free.\\[1mm]
{\bf Cons}: \\
Topological quantum matter as dark matter and its topological discrete gauge interactions with the SM
are difficult to be detected by the conventional experiments.
}} \\ 
\hline 
     \end{tabular}
\caption{Comparison of three different leptogenesis scenarios.
When we refer 
${{\bf B} -  {\bf L}}$,
we really mean ${{\bf Q} - N_c {\bf L}}$,
while its U(1) version is the  
$\U(1)_{{\bf Q} - N_c {\bf L}}$
with properly quantized charges.
We refer to
the $X$ symmetry
as a {chiral} symmetry
with $X \equiv 5({ \mathbf{B}-  \mathbf{L}})-4Y \equiv 
5({ \mathbf{B}-  \mathbf{L}})-\frac{2}{3} {\tilde Y}
=\frac{5}{N_c}({ \mathbf{Q}-  N_c \mathbf{L}})-\frac{2}{3} {\tilde Y}$ that acts chirally on the SM Weyl fermions.
We consider the energy scale $E$ with respect to
various physically pertinent energy:
Majorana mass 
$M$ and topological order gap $\Delta_{\rm TO}$, or the electroweak Higgs $\Lambda_{\rm EW}$ scale.
SSB stands for the conventional ``spontaneous symmetry breaking.'' 
}
\label{table:leptogenesis}
\end{table}

The three scenarios studied here are general and universal in the sense of
anomaly cancelation consideration,
given the systematic classification of anomalies in
Refs.~\cite{GarciaEtxebarriaMontero2018ajm1808.00009, DavighiGripaiosLohitsiri2019rcd1910.11277, WW2019fxh1910.14668}
and
\cite{JW2006.16996, JW2008.06499, JW2012.15860, WangWanYou2112.14765, WangWanYou2204.08393}. 
Regardless of which of the three (Majorana, gravitational, or topological) leptogenesis scenarios that nature may choose or not,
The three scenarios will need to welcome new experimental signatures for verification.

\newpage
\section*{Acknowledgements}

The idea of topological leptogenesis in this article has been presented by the author and circulated in his various seminars since the work of \cite{JW2012.15860} in 2020. JW thanks Yuta Hamada and Pavel Putrov
for discussions, in 2022 and 2023 respectively. 
JW also thanks Hongjian He, Nathan Seiberg, 
and Tracey Slatyer for inspiring comments in 2024.

\bibliography{BSM-Proton-Leptogenesis}

\end{document}